\documentclass[letterpaper]{aa520}
\usepackage{txfonts}
\usepackage{graphicx}
\usepackage{amssymb}
\usepackage[below]{placeins}
\usepackage{natbib}
\bibpunct{(}{)}{;}{a}{}{,}

\begin{document}

\title{Hard X-ray timing and spectral properties of \object{PSR B0540-69}}

\author{J. de Plaa \inst{1,2}
 \and L. Kuiper \inst{1}
 \and W. Hermsen \inst{1}}

\institute{SRON National Institute for Space Research, Sorbonnelaan 2, 3584 CA Utrecht, The Netherlands\\
\email{J.de.Plaa@sron.nl; L.M.Kuiper@sron.nl; W.Hermsen@sron.nl}
\and Astronomical Institute, Utrecht University, PO Box 80000, 3508 TA Utrecht, The Netherlands}
 
\offprints{J. de Plaa,\\ \email{j.de.plaa@sron.nl}}

\date{Received 19 November 2002 / Accepted 9 Januari 2003}

\abstract{
We report the hard X-ray properties of the young Crab-like LMC pulsar \object{PSR B0540-69},
using archival RXTE PCA (2 -- 60 keV) and RXTE HEXTE (15 -- 250 keV) data.
Making use of the very long effective exposure of 684 ks, we derived a very detailed 
master pulse profile for energies 2 -- 20 keV. We confirm the broad single-pulse shape
with a dip in the middle and with a significant fine structure to the left of the dip.
For the first time pulse profiles in the 10 -- 50 keV energy interval are shown.
Remarkably, the coarse pulse shape is stable from the optical up to X-ray energies 
analogous to the case of the Crab pulsar (\object{PSR B0531+21}). The profiles
can be described with two Gaussians with a phase separation of $\sim$ 0.2; the distribution
of the ratios between the two components from the optical to the X-ray range is consistent 
with being flat. Therefore we cannot conclude that the profile consists of two distinct components.
We also derived a new total pulsed spectrum in the $\sim$ 0.01 -- 50 keV range in a consistent
analysis including also archival ROSAT PSPC (0.01 -- 2.5 keV) data. This spectrum
cannot be described by a single power-law, but requires an additional energy dependent term. 
The bending of the spectrum around 10 keV resembles that of the
Crab pulsar spectrum. Although model calculations using Outer Gap scenarios could probably
explain the high-energy characteristics of PSR B0540-69 as they successfully do for the Crab, our measurements
do not entirely agree with the latest calculations by \citet{zhang-theorie}. The small discrepancies
are likely to be caused by uncertainties in the pulsar's geometry.

}

\maketitle

\section{Introduction}

\object{PSR B0540-69} is a young Crab-like pulsar located in the Large 
Magellanic Cloud (LMC) and 
was discovered at X-rays by \citet{seward} who used the Einstein X-ray 
Observatory. Like the \object{Crab} it is embedded in a bright synchrotron 
wind nebula which was confirmed in the optical waveband \citep{chanan} 
shortly after the discovery of the pulsar. The rotational characteristics of 
\object{PSR B0540-69} resemble those of the Crab pulsar 
as well, therefore their spin-down ages are of the same order of magnitude, 
$\sim 10^{3}$ yr. 

The pulse profile of \object{PSR B0540-69}, however, is considerably different from 
that of the \object{Crab}. The \object{Crab} profile shows two sharp peaks from 
optical to high-energy gamma-rays (for the most recent high-energy picture from soft 
X-rays up to high-energy gamma-rays see \citet{kuiper}), while the pulse profile of 
\object{PSR B0540-69} consists mainly of a broad asymmetric pulse with indications 
for fine structure \citep{middleditch,boyd,seward}. Most recently,
\citet{mineo} described the pulse shape as nearly sinusoidal with a minor structure on 
the left side of the maximum on the basis of higher-statistic BeppoSAX data. \citet{hirayama}
analysed all available ASCA data on this source. They confirmed the broad profile and
discussed a hump on the leading wing of the profile as a possible interpulse, analogous
to the case of the Crab.

Measurements of the total pulsed spectrum of \mbox{\object{PSR B0540-69}} 
between 0.2 and 10 keV by the Chandra X-ray Observatory suggest a hard 
spectrum with a photon spectral index of 1.83 $\pm$ 0.13 \citep{kaaret},
consistent with the BeppoSAX \citep{mineo} and ASCA \citep{hirayama} spectral
findings. At soft X-rays (ROSAT 0.1 -- 2.4 keV) \citet{finley} reported an index
1.3 $\pm$ 0.5 (90\% errors), statistically consistent with the indices measured
above 2 keV. If the pulsed spectrum would extend from medium energy X-rays with 
the same power-law index into the hard X-ray or soft gamma-ray regime, then
pulsed emission should have been detected by the BATSE (20 -- 600 keV) and OSSE 
(50 keV -- 10 MeV) instruments aboard the Compton Gamma Ray Observatory (CGRO). 
However, no detection of pulsed soft gamma-ray emission from \object{PSR B0540-69} 
has been reported sofar (\citet{hertz}; \citet{wilson}). 

The Rossi X-ray Timing Explorer (RXTE) archive contains a huge amount of hard X-ray 
data with PSR B0540-69 in the field of view. A part of these data had been used
by \citet{zhang-obs} to provide an accurate measurement of the spin frequency of 
the pulsar and its first and second derivatives by phase-linking on a limited time 
interval. 
In this paper we present the hard X-ray characteristics of PSR B0540-69 using all these
available RXTE Proportional Counter Array (PCA, 2 -- 60 keV) and RXTE High Energy X-ray 
Timing Experiment (HEXTE, 15 -- 250 keV) data. In order to augment the
spectral coverage towards lower energies we also used soft X-ray data ($\sim$ 0.01 -- 2.5 keV)
from the ROSAT PSPC in a consistent analysis.

\section{Observations and data}

\subsection{RXTE PCA}

The PCA \citep{pca} consists of five Proportional Counter Units 
(PCU) sensitive in the $\sim$2 -- 60 keV range with an energy resolution less than 
18\% at 6 keV and a large field of view of 1$\degr$ FWHM. 
Each PCU has three Xenon layers which provide the basic scientific 
data. When running in GoodXenon configuration each photon captured in a Xenon
layer is registered with a timing resolution of 1 $\mu$s.
The individual
triggers do not contain any spatial information.

The PCA observed the LMC many times with PSR B0540-69 in its large field of view.
We analysed archival
data from observations performed in the GoodXenon configuration with 
offset angles less than 30${\arcmin}$ and durations longer than $\sim$5 ks (see Table~\ref{tab:all_obs}), 
resulting in a total dataset containing 684 ks of exposure, after screening 
for Earth occultations, South Atlantic Anomaly (SAA) passages and enhanced 
background due to high-energy particles. We used this exceptionally long exposure of 
PSR B0540-69 for our timing analysis. To boost the signal-to-noise ratio for this weak 
source we selected the 
photons registered in the first Xenon layer of the PCUs.

\begin{table}
\begin{center}
\caption{List of all RXTE observations used in this study with PSR B0540-69
 within 30$\arcmin$ of the pointing direction. The exposures mentioned are effective 
 values screened for Earth occultations, SAA passages and contaminating particles.}
\begin{tabular}{llrr}
\hline\hline
Obs.ID & Target & Offset & Exposure\\
        &        & ($\arcmin$) & (ks)\\
\hline
10206 & \object{PSR B0540-69} & 0.04  & 39.1\\
10218 & \object{SN1987A}      & 25.30 & 74.1\\
10250 & \object{LMC X-1}      & 24.80 & 33.4\\
20188 & \object{LMC X-1}      & 24.80 & 234.8\\
30087 & \object{LMC X-1}      & 24.77 & 73.5\\
40139 & \object{PSR J0537-69} & 15.96 & 109.0\\
50103 & \object{PSR J0537-69} & 15.94 & 120.0\\
\hline
\end{tabular}
\label{tab:all_obs}
\end{center}
\end{table}

\subsection{RXTE HEXTE}

The HEXTE instrument \citep{rothschild} aboard RXTE consists of two independent detector 
clusters (A and B) each containing 4 Na(Tl)/CsI(Na) phoswich scintillation detectors passively
collimated to a 1$\degr$ FWHM field of view and co-aligned with the PCA.
The instrument is sensitive to photons with energies in the energy range 
15 -- 250 keV with an energy resolution of about 15.4\% at 60 keV. The net open
area of the 8 detectors is $1600$ cm$^2$. The events can be tagged with a
maximum time resolution of $7.6\mu$s. In its default operation mode the field of
view of each cluster is switched on and off source to provide instantaneous background 
measurements reducing the effective source-on exposure by roughly
a half. Due to high-energy particle events the overall performance of
the instrument is degraded resulting in a dead-time fraction of about 40\%.
When the source of interest is observed off-axis the response is further reduced by the 
collimator/detector assembly. For a source observed $30\arcmin$ off-axis the sensitive 
area is reduced to about 0.5 of its on-axis value.

In the spectral analysis of HEXTE data all these effects which reduce the effective
sensitive area are taken into account. Due to the co-alignment of HEXTE with 
the PCA the same observations have been used as for the PCA. For the spectral
analysis HEXTE data from all observations listed in Table~\ref{tab:all_obs} have
been used, not limited to those observations falling within the stable gain 
period used in the PCA spectral analysis. 

The total dead-time and off-axis corrected Cluster-A and B on-source (=PSR B0540-69) 
exposures are 248.4 ks and 260.7 ks, respectively.

\subsection{ROSAT PSPC}
\label{sec:rosat}

To extend the spectral coverage to the soft X-ray regime we also included ROSAT PSPC
data in our study. The PSPC is a gas-filled imaging proportional counter which operates
in photon counting mode \citep{Pfeffermann}. It is sensitive over the 0.01 -- 2.5 keV 
energy band. The energy resolution is $\Delta E=0.45 E^{0.5}$. The events are tagged
on board with a precision of $130\mu$s, but the erratic spacecraft clock makes absolute
timing difficult.  

We could identify in the ROSAT archive 3 PSPC observation sequences with PSR B0540-69 
nearly on-axis. The observation identifiers are RP400052N00 (PSPC-B), 
RP150044N00 (PSPC-C) and RP400133N00 (PSPC-B), with effective exposure times of 8.5 ks,
5.1 ks and 1.7 ks, respectively.

\section{Timing analysis}

For all selected photons measured with the RXTE PCA with energies above 2 keV barycentric corrections 
to their arrival times have been derived using the FTOOLS program 
\verb+fasebin+ \citep{ftools} and a source position of $\alpha_{2000}$ = 05$^{h}$40$^{m}$11$^{s}$.221 and $\delta_{2000}$
= -69{\degr}19{\arcmin}54{\arcsec}.98 \citep{kaaret}. The ephemeris used 
to derive the pulse profile is presented in Table~\ref{tab:ephemeris}. 
This ephemeris predicts accurately the instantaneous rotation frequency
of the source for the epochs of our observations. However, it is not a phase connected 
timing solution.

\begin{table}
\caption{Ephemeris of PSR B0540-69 determined by \citet{kaaret}.}
\begin{center}
\begin{tabular}{ll}
\hline\hline
Parameter & Value \\
\hline
Val. range (MJD) & 46\,992 -- 51\,421 \\
$t_{0}$ (MJD) & 47\,700.0 \\
$\nu_{0}$ (Hz) & 19.8593584982(40) \\
$\dot{\nu}_{0}$ ($10^{-10}$  Hz s$^{-1}$) & -1.8894081(7) \\
$\ddot{\nu}_{0}$ ($10^{-21}$ Hz s$^{-2}$) & 3.7425(43) \\
$\alpha_{2000}$ & 05$^{h}$40$^{m}$11$^{s}$.221 \\
$\delta_{2000}$ & -69{\degr}19{\arcmin}54{\arcsec}.98 \\
\hline
\multicolumn{2}{c}{Note: Numbers in parentheses are} \\
\multicolumn{2}{c}{1$\sigma$ errors in the last quoted digits.} \\
\end{tabular}
\end{center}
\label{tab:ephemeris}
\end{table}

For each observation contributing to our total data set of 684 ks we 
derived a pulse profile. Because the ephemeris is not a phase connected solution for 
long timescales, we cannot combine the individual profiles directly. Therefore, we correlated all 
pulse profiles first with 
a profile template based on the longest data block and later again with the result of the 
aligned combination to obtain the final profile. The statistical uncertainties in the
estimates of the phase shifts varied approximately over the range 0.6 -- 3\%.

\begin{figure}
\resizebox{\hsize}{!}{\includegraphics{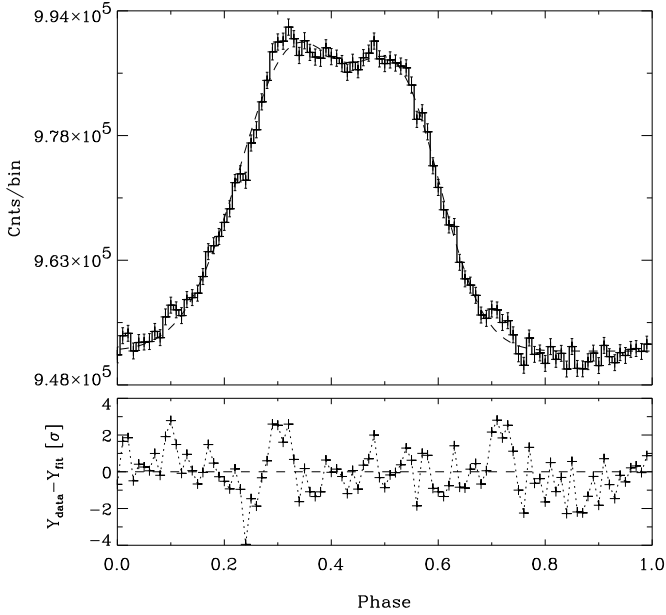}}
\caption{Master pulse profile of PSR B0540-69 derived from the total 684 ks RXTE PCA data 
set for energies 2 -- 20 keV. The broken line shows the best fit empirical model consisting
of two symmetric Gaussian profiles separated $\sim$0.2 in phase. The lower figure shows the distribution 
of the residuals in number of $\sigma$'s.}
\label{fig:profile_max_stat}
\end{figure}

Fig.~\ref{fig:profile_max_stat} shows the resulting pulse profile (2 -- 20 keV) of PSR B0540-69
for the total screened exposure of 684 ks. The statistical accuracy of this profile is at least an 
order of magnitude higher than for earlier published profiles (e.g. \citet{eikenberry}, 
\citet{mineo}, \citet{hirayama}) and the genuine profile becomes more clearly visible: 
A single broad pulse with a clear dip at the top. Such a shape could be the result of 
two narrower pulses separated $\sim$ 0.2 in phase. To investigate this further, we
attempted to describe the shape of the profile as the sum of two symmetrical Gaussians on top of a flat background, by
making a fit to the measured profile with as free parameters the widths, positions and amplitudes 
of the gaussians. The resulting best fit is shown in Fig.~\ref{fig:profile_max_stat}
(positions: $p_{\mathrm{1}}$ = 0.330 $\pm$ 0.004, $p_{\mathrm{2}}$ = 0.531 $\pm$ 0.003; 
widths: $\sigma_{\mathrm{1}}$ = 0.102 $\pm$ 0.003, $\sigma_{\mathrm{2}}$ = 0.079 $\pm$ 0.002) 
with underneath the fit residuals expressed 
in number of $\sigma$'s.  The overall fit looks good, particularly for the component with 
maximum at phase 0.531, but there are a few phase intervals in which there appear significant  
deviations from this empirical model ($\chi^2_r$ of fit: 167.3/93 ; $P(\chi^2 > 167.3 \mid 93)$ = 3.65 $\times$ 10$^{-6}$ (4.7$\sigma$ dev.)). 
The most significant positive excess is around phase 0.3
(near the position of one of the gaussians)
reaching a significance of 5.4$\sigma$, proving that the top of the broad profile
has significant fine structure. The positive excesses far in the wings at
phases around 0.1 and 0.7 reach significances of 3.6 and 4.0$\sigma$, respectively.
We do not consider these latter deviations significant, because there was no a priori reason to
select these phases (no single trial). \citet{hirayama} discussed a hump in the ASCA
profile at a phase consistent with our excess around phase 0.1, as a possible separate 
interpulse in analogy to the Crab profile. With our better statistics we do not confirm
the detection of this hump. Finally, the negative excursion at phase 0.24, together with 
the positive excess at phase 0.3 suggest that the leading wing of the profile is steeper than
fitted with the Gaussian profile.

\begin{figure*}
\begin{center}
 \includegraphics[width=1.0\textwidth]{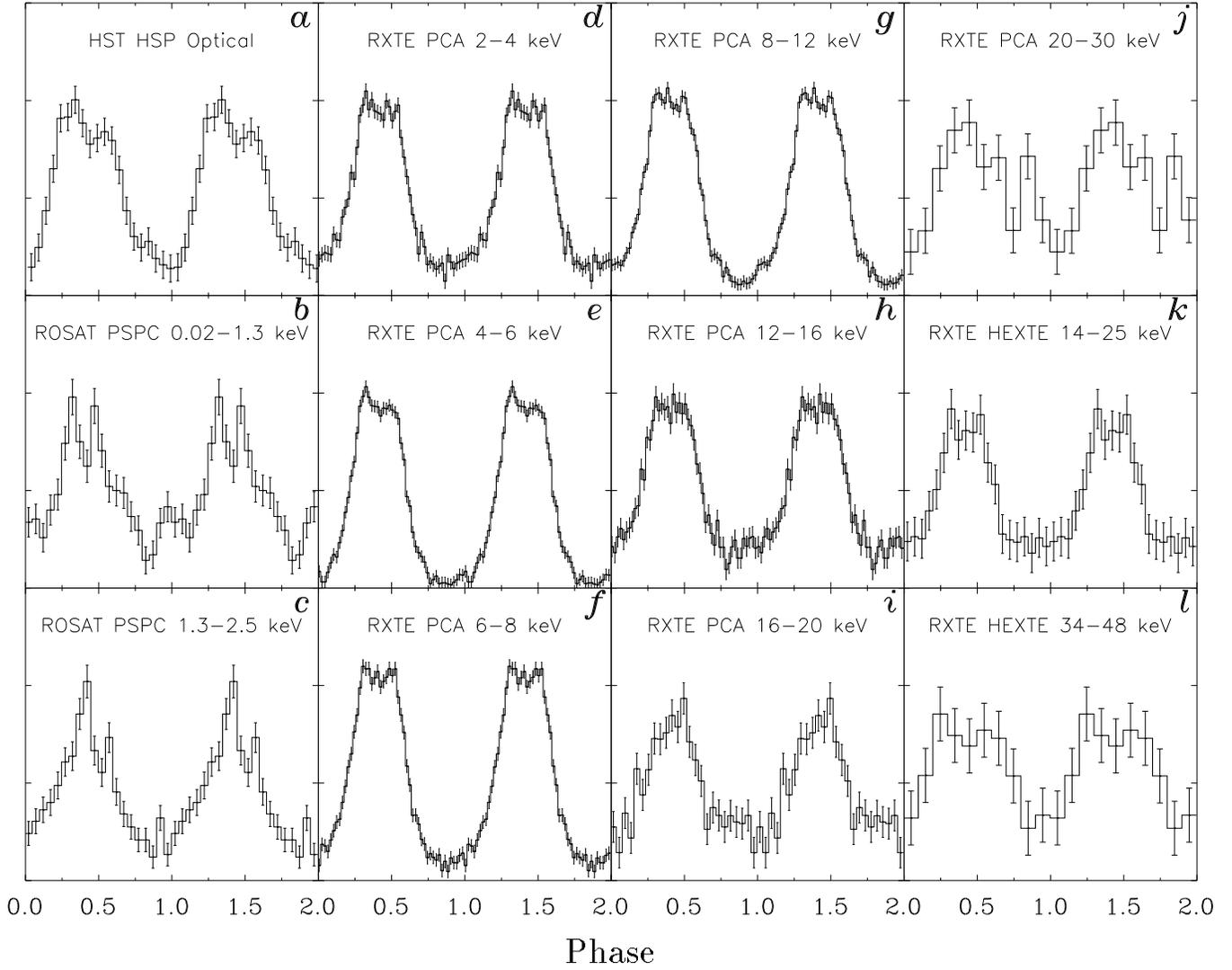}
\end{center}
 \caption{Pulse profiles of PSR B0540-69 from the optical window to hard X-rays. \textbf{a}) Optical profile 
 as observed by \citet{boyd} with Hubble's High Speed Photometer (HSP). \textbf{b-c}) ROSAT PSPC,
 energy ranges 0.02--1.3 and 1.3--2.5 keV. \textbf{d-j}) RXTE PCA, energy ranges 2--4, 4--6,
 6--8, 8--12, 12--16, 16--20 and 20--30 keV. \textbf{k-l}) RXTE HEXTE, energy ranges 14--25 and 
 34--48 keV. The RXTE HEXTE light\-curve for energies 25 -- 34 keV is not shown since it does not 
give a significant signal due to high instrumental background. 
The ROSAT and HST profiles have been aligned in phase to the RXTE profiles by cross correlation.}
\label{fig:all_lightcurves}
\end{figure*}

The high counting statistics in the PCA data allow us to construct significant pulse profiles in 
differential energy windows from 2 keV up to 30 keV (Fig.~\ref{fig:all_lightcurves} {\it d} to {\it j}),
showing the first profiles above 10 keV. Up to 12 keV the statistical accuracy is very high and some 
fine structure on top of the broad profile seems to be consistently present in successive energy windows, 
most notably the sharp maximum at phase $\sim 0.3$.
 
Fig.~\ref{fig:all_lightcurves} also shows pulse profiles we
derived from RXTE HEXTE data at higher energies and from ROSAT data at lower energies.
We could construct the HEXTE profiles of \object{PSR B0540-69} by following the same analysis steps
for the same observation windows as was performed for the PCA timing analysis, applying the phase shifts
determined in the correlation analysis of the high-statistics PCA profiles to the low-statistics
HEXTE profiles. This allowed us to reveal the \object{PSR B0540-69} pulse profile up to 48 keV.
The ROSAT profiles are derived using the data reported earlier in Sect.~\ref{sec:rosat}, and the phase 
alignment was determined by
cross correlating the PCA master profile with the total ROSAT profile. Finally, the optical \object{PSR B0540-69}
pulse profile from the High Speed Photometer (HSP) aboard the Hubble Space Telescope (HST) \citep{boyd} is 
shown. The optical profile shape is strikingly similar to the PCA X-ray profile. Therefore, we choose to 
align also the optical profile by cross correlation. We note, however, that \citet{ulmer} report in a 
preliminary communication that there is a phase shift of 0.17 between the optical and the X-ray pulse, 
the optical arriving earlier. We would like to see this result confirmed.

\begin{figure}
\resizebox{\hsize}{!}{\includegraphics{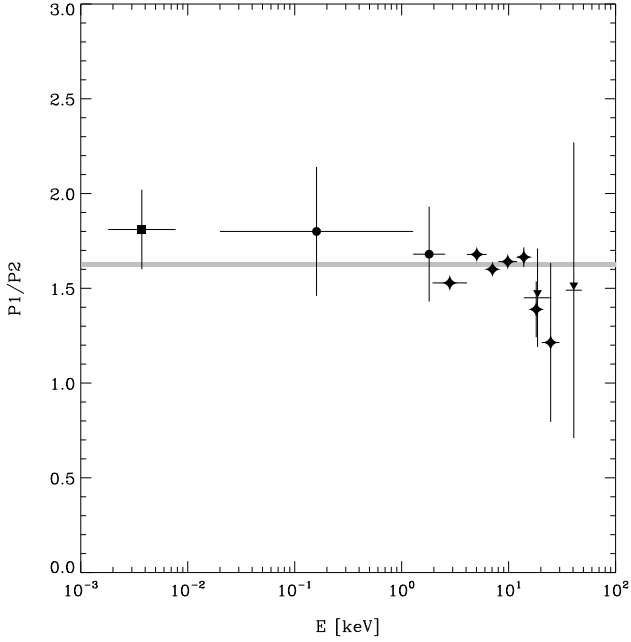}}
\caption{Ratio of the pulsed (excess) counts assigned to the two Gaussians components of the model fit to the
profiles shown in Fig.~\ref{fig:all_lightcurves}. This distribution is consistent with being flat with
a P1/P2 value of 1.63 $\pm$ 0.01 with a $\chi^2_r$ of 21.66 / 11 resulting in a random probability of 2.7\%.
Data points are derived from the optical HST HSP (square), ROSAT PSPC (circle), RXTE PCA (diamond) and RXTE HEXTE (triangle) observations.}
\label{fig:ratio_gauss}
\end{figure}

At a first glance it is not obvious, due to the varying statistics, whether the pulse shape varies 
from the optical to the hard X-ray windows.
Therefore, we fitted now all profiles in Fig.~\ref{fig:all_lightcurves} with the empirical model of
two Gaussian profiles with the positions and the widths fixed to the best-fit values shown in 
Fig.~\ref{fig:profile_max_stat}
and leaving the normalisations free. We verified that all fits are good with reduced $\chi^2$ values ranging between
0.93 and 1.32 (98 d.o.f), except for the optical profile (18 d.o.f.). Only this optical profile appears to be 
somewhat broader than the X-ray profiles. 
Fig.~\ref{fig:ratio_gauss} shows the ratio of the counts in the two Gaussian components
from the optical regime to the hard X-rays. This ratio appears to be surprisingly constant, underlining the stability 
of the profile shape and showing that the spectra of the two empirical components are within the statistics the same.
Therefore, we cannot conclude that the broad and structured pulse profile of \object{PSR B0540-69} is the sum of two
distinct components with different spectra.

\section{Spectral analysis}
\label{sec:data_spec}

Unfortunately, the total RXTE PCA data set of 684 ks is not uniform in energy calibration, because
the gain settings on the PCA instrument have been changed a number of times 
since the launch of RXTE. For the derivation of the total pulsed spectrum of PSR B0540-69
we selected therefore the observations performed in the third and longest 
gain epoch (April 15, 1996, 23:05 -- March 22, 1999, 17:37), to obtain a dataset 
with a constant channel-to-energy relation with total exposure of 462 ks. For each observation
in this data set we derived the pulse profile and the response corrected for the offset angle. 
If an observation
contained one or more time intervals in which the number or the order 
of active PCUs was different, then these intervals were treated as separate 
data blocks. All offset-corrected 
response files were summed using weighing factors 
($f_{i}$) as calculated in Eq.~(\ref{eqn:factor}), where $N_{\mathrm{pcu},i}$ 
is the number of PCUs on in data block $i$, $t_{i}$ is the exposure of data block $i$ 
and $T$ is the total exposure of the selected (third epoch) dataset. 

\begin{equation}
f_{i} = \frac{N_{\mathrm{pcu},i} t_{i}}{5 T} 
\label{eqn:factor}
\end{equation}

Pulse profiles in 14 differential energy ranges between 3.4 and 30.2 keV for the 462 ks data set were 
derived following the same procedure as explained 
in the timing analysis. 

In order to determine the excess counts in the broad pulse profiles for all
14 differential energy ranges, we fitted these profiles with the shape of the best-model fit
shown in Fig.~\ref{fig:profile_max_stat}. Again, we verified that the model
fits were statistically in agreement with the data for all energy intervals.
Using the numbers of excess counts and the derived response values we can derive the PSR B0540-69 pulsed spectrum
measured by the PCA.

\begin{figure}
\resizebox{\hsize}{!}{\includegraphics{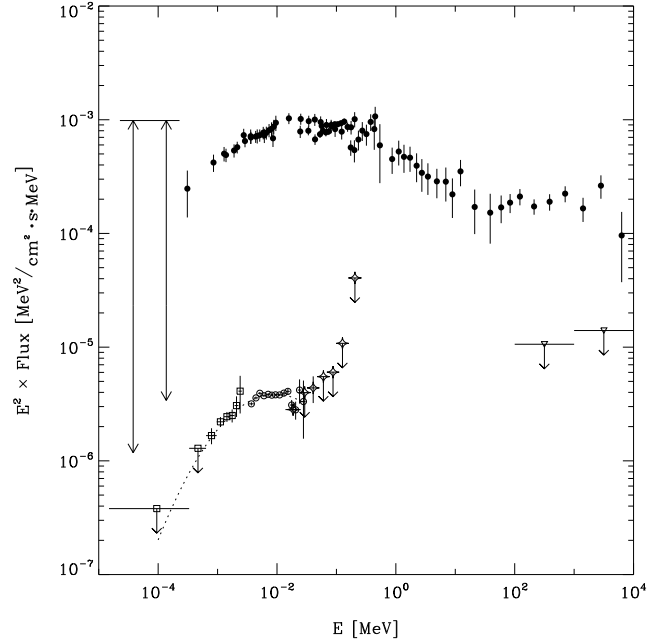}}
\caption{Total pulsed spectrum of the Crab pulsar and \object{PSR B0540-69}  
 for X-ray energies above 0.01 keV. Crab (\textit{top}): Flux measurements taken from \citet{kuiper}. 
 PSR B0540-69 (\textit{bottom}): Flux measurements
 and upper limits from ROSAT PSPC (open squares, 0.01--2.5 keV), RXTE PCA (open circles, 3.4--30 
 keV) and RXTE HEXTE (open diamonds, 14--300 keV) are shown, as well as upper limits at 
 high-energy $\gamma$-rays from CGRO EGRET \citep{thompson}. 
 The sizes of the arrows on the left indicate the shift in flux ($\pm1\sigma$) to put 
 the Crab pulsar at the LMC distance. 
 We used 2.0 $\pm$ 0.5 kpc for the Crab and 49.4 $\pm$ 3.4 kpc 
 for the \object{PSR B0540-69} (LMC) distance.}
\label{fig:crab_comparison}
\end{figure}

\begin{table}
\caption{Flux values derived from the ROSAT PSPC, RXTE PCA and RXTE HEXTE data 
(see Fig.~\ref{fig:crab_comparison}). In this table we give 2$\sigma$ upper limits.}
\begin{center}
\begin{tabular}{r @{ -- } l r}
\hline\hline
\multicolumn{2}{c}{Energy interval} & \multicolumn{1}{c}{Flux} \\
\multicolumn{2}{c}{(keV)} & \multicolumn{1}{c}{(photons s$^{-1}$ cm$^{-2}$ keV$^{-1}$)} \\
\hline
\multicolumn{3}{c}{ROSAT PSPC}  \\
\hline
0.01 & 0.33 &  $<$ 3   $\times$ 10$^{-2}$ \\
0.33 & 0.63 &  $<$ 5.8 $\times$ 10$^{-3}$ \\
0.64 & 0.97 &  (2.6 $\pm$ 0.4) $\times$ 10$^{-3}$ \\
0.97 & 1.28 &  (1.8 $\pm$ 0.2) $\times$ 10$^{-3}$ \\
1.28 & 1.60 &  (1.2 $\pm$ 0.1) $\times$ 10$^{-3}$ \\
1.60 & 1.93 &  (8.0 $\pm$ 1.0) $\times$ 10$^{-4}$ \\
1.93 & 2.25 &  (7.0 $\pm$ 1.4) $\times$ 10$^{-4}$ \\
2.25 & 2.55 &  (7.1 $\pm$ 2.6) $\times$ 10$^{-4}$ \\
\hline
\multicolumn{3}{c}{RXTE PCA}  \\
\hline
3.37 & 4.08 & (2.31 $\pm$ 0.07) $\times$ 10$^{-4}$ \\
4.08 & 4.79 & (1.83 $\pm$ 0.05) $\times$ 10$^{-4}$ \\
4.79 & 5.51 & (1.49 $\pm$ 0.04) $\times$ 10$^{-4}$ \\
5.51 & 6.58 & (1.03 $\pm$ 0.02) $\times$ 10$^{-4}$ \\
6.58 & 7.66 & (7.63 $\pm$ 0.15) $\times$ 10$^{-5}$ \\
7.66 & 8.74 & (5.65 $\pm$ 0.12) $\times$ 10$^{-5}$ \\
8.74 & 10.2 & (4.28 $\pm$ 0.09) $\times$ 10$^{-5}$ \\
10.2 & 12.0 & (3.11 $\pm$ 0.07) $\times$ 10$^{-5}$ \\
12.0 & 14.2 & (2.32 $\pm$ 0.08) $\times$ 10$^{-5}$ \\
14.2 & 16.4 & (1.76 $\pm$ 0.09) $\times$ 10$^{-5}$ \\
16.4 & 19.0 & (10.0 $\pm$ 0.1) $\times$ 10$^{-5}$ \\
19.0 & 22.3 & (6.6 $\pm$ 1.2) $\times$ 10$^{-6}$ \\									      
22.3 & 25.7 & (7.3 $\pm$ 1.8) $\times$ 10$^{-6}$ \\
25.7 & 30.2 & (4.3 $\pm$ 2.3) $\times$ 10$^{-6}$ \\
\hline
\multicolumn{3}{c}{RXTE HEXTE}  \\			
\hline							    
13.9 & 25.1 & (8.1 $\pm$ 0.7) $\times$ 10$^{-6}$ \\
25.1 & 34.2 & (1.8 $\pm$ 1.4) $\times$ 10$^{-6}$ \\
34.2 & 48.4 & (2.6 $\pm$ 0.7) $\times$ 10$^{-6}$ \\
48.4 & 75.4 & $<$ 1.5 $\times$ 10$^{-6}$ \\
75.4 & 102. & $<$ 7.8 $\times$ 10$^{-7}$ \\
102. & 157. & $<$ 6.8 $\times$ 10$^{-7}$ \\
157. & 268. & $<$ 9.6 $\times$ 10$^{-7}$ \\
\hline
\end{tabular}
\end{center}
\label{tab:flux_values}
\end{table}

For the ROSAT and RXTE HEXTE data we performed a consistent spectral analysis. Also for the profiles
measured by these instruments we determined the pulsed excess counts by making fits with the 
empirical model. The resulting total pulsed X-ray spectrum  of PSR B0540-69 is shown in 
Fig.~\ref{fig:crab_comparison}; the flux values are given in Table~\ref{tab:flux_values}. 
From the plot it is evident that the X-ray spectrum has not the shape of a single power-law. 
A single absorbed power-law fit to just the ROSAT PSPC and RXTE PCA data renders a photon 
index of 1.845 $\pm$ 0.004, fixing the hydrogen column density 
(N$_{\mathrm{H}}$) at 4.6 $\times$ 10$^{21}$ cm$^{-2}$ \citep{kaaret}. 
This value is consistent with earlier measurements: 1.83 $\pm$ 0.13 \citep{kaaret}, 
1.94 $\pm$ 0.03 \citep{mineo} and 1.3 $\pm$ 0.5 \citep{finley}. However, our reduced $\chi^2$ 
is unacceptably high with a value of 4.75 with 20 d.o.f., rejecting this simple shape.
Therefore we have fitted the data with an absorbed power-law model with an energy dependent index:
\begin{equation}
F(E_{\gamma}) = e^{-\mathrm{N}_{\mathrm{H}} \cdot \sigma} \alpha E_{\gamma}^{-(\beta + \gamma \ln E_{\gamma})} 
\label{eqn:epowerlaw}
\end{equation}
In the fit the parameters $\alpha$, $\beta$ and $\gamma$ were kept free and
N$_{\mathrm{H}}$ was fixed again at $4.6 \times 10^{21}$ cm$^{-2}$. 

This resulted in a $\beta$ of 1.360 $\pm$ 0.005
and $\gamma$ of 0.143 $\pm$ 0.003, which
means that the photon index softens at higher energies (for example at 1 keV : $\sim$ 1.35, 2 keV : $\sim$ 1.56,
4 keV : $\sim$ 1.76 and 8 keV : $\sim$ 1.95).
With a $\chi^2_r$ of 31.0/19
the model fits reasonably well to the measured spectrum. For energies 
between 20 and 50 keV, the measured shape
is poorly defined, and the
flux values of the PCA and HEXTE are also consistent with a power-law with index 2.

\section{Summary and discussion}
 
This study of the hard X-ray characteristics of the young spin-down powered pulsar \object{PSR B0540-69} 
using archival RXTE PCA/HEXTE and ROSAT PSPC
observations resulted in a detection of this pulsar up to $\sim$ 50 keV. For the first time pulse profiles
above 10 keV are shown. Furthermore, we derived the total pulsed spectrum of PSR B0540-69 in a consistent 
analysis from $\sim$ 0.01 -- 50 keV.

Because of the high statistics we were able to obtain a very detailed master pulse 
profile in the 2 -- 20 keV range, which could reasonably be described by the sum of two symmetrical Gaussians. 
Remarkably, this double Gaussian shape could be well fitted to all shown X-ray pulse profiles and 
resembles very much the optical 
profile which is, however, somewhat broader. The distribution of the ratios between the two Gaussian 
components as a function of energy appears to be consistent with being flat. Therefore we could not conclude 
that the pulse consists of two distinct Gaussian components with different spectra.

The derived total pulsed spectrum of PSR B0540-69 clearly shows a bend towards higher energies
and cannot be described by a single power-law. We proposed and fitted an energy dependent 
power-law model to the data, successfully describing the curved spectral shape up to $\sim$ 20 keV. 
Between 20 and 50 keV the measured flux shape is also consistent with a power-law with index 2.

Directly after the discovery of PSR B0540-69 its timing properties, apart from the pulse profile, were found 
to be very Crab-like \citep{seward}. If we compare our results to the characteristics of the Crab pulsar 
we see some remarkable additional
similarities. Even though the Crab and PSR B0540-69 show different pulse profiles (Crab: two sharp peaks,
PSR B0540-69: one broad structured pulse), for both the coarse shapes of the profiles remain stable 
from optical to hard X-rays. The second Crab-like pulsar (\object{PSR B1509-58}) also shows
a very stable single broad pulse from soft X-rays up to MeV $\gamma$-rays \citep{kuiper1999,cusumano},
but pulsed optical emission has not been found sofar.

Furthermore, the shape of the total pulsed spectrum of PSR B0540-69 looks strikingly similar to that of the Crab   
in the $\sim$ 0.01 -- 50 keV range. In Fig.~\ref{fig:crab_comparison} the two total pulsed spectra are plotted in E$^2$F(E)
representation. The bending of the spectrum around 10 keV appears to occur similarly in both spectra.
The intrinsic X-ray luminosities are comparable as well, within the uncertainties caused by the errors in
the Crab distance and the absolute calibrations of the RXTE PCA and HEXTE detectors.

Next to PSR B0540-69 and the Crab pulsar, \object{PSR B1509-58} shows a similar spectral behaviour at X-ray/$\gamma$-ray
energies \citep{kuiper1999,cusumano}, although its spectrum appears to reach its maximum
luminosity at higher energies (10 -- 30 MeV) relative to the Crab and PSR B0540-69. These three young 
($\lesssim 1.6 \times 10^{3}$ yr) X-ray/$\gamma$-ray pulsars exhibit different spectral behaviours compared to older $\gamma$-ray pulsars 
like, for example, the \object{Vela} pulsar and \object{Geminga}. Older $\gamma$-ray pulsars show a hard high-energy spectrum with a 
turn-over at GeV energies. Therefore, the observed spectral shape for the three young pulsars might be 
characteristic for very young pulsars in general.

The models that have been proposed to explain high-energy emission from highly magnetised pulsars, can be
classified into two distinct classes, namely Polar Cap (e.g. \citet{sturrock,daugherty}) and Outer Gap models
(e.g. \citet{cheng_1986,cheng_1986_2,chiang,romani,cheng}). Outer gap models seem to be
most successful in explaining the Crab pulse profile and spectrum as well as the optical polarisation angle
variation with phase \citep{romani,cheng}. 

As PSR B0540-69 shows predominantly Crab-like
characteristics, one may expect that Outer Gap models can explain its profile and spectrum as well.
Unfortunately, the geometry of the magnetic inclination ($\alpha$) and viewing angles ($\zeta$) cannot be constrained by radio 
measurements. In a recent model calculation for PSR B0540-69 based on the three dimensional outer magnetosphere
model of \citet{cheng}, \citet{zhang-theorie} chose likely values for 
these angles ($\alpha = 50\degr$; $\zeta = 76\degr$)
by constraining the model to the observed radio emission and broad pulse profiles. Their
calculations predict a single broad pulse shape in the optical, X-ray and $\gamma$-ray windows and a power-law shaped 
total pulsed spectrum from the optical up to $\sim$ 100 keV. Above 1 MeV a bend in the spectrum is expected.

Our new hard X-ray findings do not agree entirely with these model calculations. The first difference is in the
observed high-resolution pulse profile. We confirmed the broad nature of the pulse, but it also shows a significant 
dip on the top of the profile with a significant excess to the left of it.
Secondly, the observed total pulsed spectrum flattens above 10 keV, while the model
expects the bend to occur above 100 keV. These discrepancies are likely a result of the uncertainties in the
pulsar geometry.  
We note that \citet{zhang-theorie} also describe with their Outer Gap model the high energy properties
of \object{PSR B1509-58}. For slightly different angles ($\alpha = 60\degr$; $\zeta = 75\degr$) they
predict a broad pulse profile with fine structure over the top, very similar to what we found for PSR B0540-69.
This suggests that model calculations for PSR B0540-69 with a slightly larger magnetic inclination angle
could reproduce in detail our measured profile.
Therefore, optical polarisation measurements are required to constrain the magnetic inclination and
viewing angles to model the high-energy emission processes of this young pulsar more accurately.

Earlier, \citet{cheng_wei} calculated the high-energy spectrum based on the Outer Gap model of
\citet{cheng_1986,cheng_1986_2} assuming that the X-ray and $\gamma$-ray emission is synchrotron
self-Compton radiation from secondary e$^{\pm}$ pairs. They predict a spectral break between
about 10 and 100 keV depending on the size of the Outer Gap. A relatively large Outer Gap
(gradually increasing with age) could explain the apparent break in the spectrum around 10 keV.

A more accurate measurement at higher energies of the total pulsed spectrum of \object{PSR B0540-69} is
very important to discriminate the different model calculations. For the INTEGRAL mission (20 keV -- 8 MeV)
a very long exposure (2 Ms) of the LMC is scheduled. Unfortunately, for the spectral shape we derived
in this work INTEGRAL will be able to detect the pulsed signal only up to $\sim$ 70 keV.  
More importantly, GLAST (20 MeV -- 300 GeV) should be able to detect pulsed hard $\gamma$-ray emission 
if the spectrum behaves Crab-like.

\begin{acknowledgements}
We wish to thank the High Energy As\-tro\-physics Science Archive Research Center (HEASARC) at NASA/Goddard
Space Flight Center for maintaining its online archive service which provided the data 
used in this research. We also thank Frank Verbunt (Utrecht University)
for carefully reading our manuscript.
\end{acknowledgements}


\nocite{xte}

\bibliographystyle{aa520}
\bibliography{pkh2003}

\end{document}